\documentclass[12pt,preprint,a4wide]{aastex}

\newcommand{\kte}{kT$_{\rm e}$}
\newcommand{\fhard}{F$_{\rm Hard}$}
\newcommand{\ktbb}{kT$_{\rm BB}$}

\newcommand{\ergs}{ergs s$^{-1}$}
\newcommand{\ergscm}{ergs s$^{-1}$ cm$^{-2}$}

\newcommand{\rxte}{{\it RXTE}}

\newcommand{\nulf}{$\nu_{\rm LF}$}
\newcommand{\nub}{$\nu_{b}$}
\newcommand{\nul}{$\nu_{l}$}

\slugcomment{In preparation for ApJ, version of \today}

\shorttitle{Correlated timing and spectral behavior of \fouru}
\shortauthors{Barret \& Olive.}

\def\fouru{4U 1705--44}
\def\4U1705{4U $1705-44$}

\begin{document}

\title{Correlated timing and spectral behavior of \fouru}

\author{Jean-Fran\c{c}ois Olive\altaffilmark{1}, Didier 
Barret\altaffilmark{1} \& Marek Gierli{\'n}ski\altaffilmark{2,3}}

\altaffiltext{1}{Centre d'Etude Spatiale des Rayonnements, CNRS/UPS, 
9 Avenue du Colonel Roche, 31028 Toulouse Cedex 04, France}

\altaffiltext{2}{Department of Physics, University 
of Durham, South Road, Durham DH1 3LE, UK}

\altaffiltext{3}{Uniwersytet Jagiello{\'n}ski, Obserwatorium 
Astronomiczne, ul. Orla 171, 30-244 Krak{\'o}w, Poland}

\email{Olive@cesr.fr}
%
%
\begin{abstract}

We follow the timing properties of the neutron star low-mass X-ray 
binary system \fouru~in different spectral states, as monitored by 
the Rossi X-ray Timing Explorer over about a month. We fit the power 
density spectra using multiple Lorentzians. We show that the 
characteristic frequencies of these Lorentzians, when properly 
identified, fit within the correlations previously reported. The time 
evolution of these frequencies and their relation with the parameters 
of the energy spectra reported in Barret \& Olive (2002) are used to 
constrain the accretion geometry changes. The spectral data were 
fitted by the sum of a blackbody and a Comptonized component and were 
interpreted in the framework of a truncated accretion disk geometry, 
with a varying truncation radius. If one assumes that the 
characteristic frequencies of the Lorentzians are some measure of 
this truncation radius, as in most theoretical models, then the 
timing data presented here strengthen the above interpretation. The 
soft to hard and hard to soft transitions are clearly associated with 
the disk receding from and approaching the neutron star respectively. 
During the transitions, correlations are found between the Lorentzian 
frequencies and the flux and temperature of the blackbody, which is 
thus likely to be coming from the disk. On the other hand, in the 
hard state, the characteristic Lorentzians frequencies which are at 
the lowest, remained nearly constant despite significant evolution of 
the spectra parameters. The disk no longer contributes to the X-ray 
emission, and the blackbody is now likely to be emitted by the 
neutron star surface which is providing the seed photons for the 
Comptonization. \end{abstract}

\keywords{X-rays: star, stars: individual: \fouru, stars: neutron,
accretion, accretion disks}

\section{Introduction}

Rapid X-ray variability is a powerful probe of physics of accretion 
flows around neutron stars and black holes. Significant progress has 
been accomplished recently with the advent of the {\it Rossi X-ray 
Timing Explorer} (\rxte, Bradt, Rothschild \& Swank 1993). Its large 
collecting area and micro-second time resolution allowed us to study 
with exquisite details the rapid variability of a wide variety of 
accreting X-ray sources. In particular, the power density spectra 
(PDS) of Galactic black hole candidates and neutron star low-mass 
X-ray binaries (LMXBs) exhibit a variety of features ranging from 
narrow quasi-periodic oscillations (QPOs) to broad noise components 
(for a review see Wijnands 2001 and references therein). 

It has been shown recently that these PDS can be well represented as 
a superposition of a few Lorentzians (e.g.\ Olive et al.\ 1998; Van 
Straaten et al.\ 2002; Belloni, Psaltis \& Van der Klis 2002). The 
multi-Lorentzian approach gives a simple and universal 
phenomenological description of the PDS. Each noise component 
(Lorentzian) is then described by its characteristic frequency, width 
and amplitude, usually expressed as a fractional root-mean-square 
(RMS). It has been demonstrated that these characteristic frequencies 
are correlated for a given source and between sources, both for 
neutron star and black hole binaries (Wijnands \& Van der Klis 1999; 
Psaltis, Belloni \& van der Klis 1999; Belloni et al.\ 2002). In most 
models of rapid X-ray variability, these frequencies are related to 
some kind of an abrupt transition in the accretion disk (e.g.\ 
Stella, Vietri \& Morsink 1999; Titarchuk \& Osherovitch 1999; 
Psaltis \& Norman 2002). This could be a truncation of a 
Shakura-Sunyaev disk into a hot inner flow (e.g.\ 
R{\'o}{\.z}a{\'n}ska \& Czerny 2000). With the truncation radius 
decreasing (the disk coming in) the characteristic frequencies 
increase.

\fouru~is a neutron star LMXB system which belongs to the class of 
atoll sources (Hasinger \& Van der Klis 1989). The source shows 
variability on all time scales, from months down to milliseconds 
(Langmeier et al.\ 1987; Berger \& Van der Klis 1998, Ford, van der 
Klis \& Kaaret 1998, Liu, van Paradijs \& van den Heuvel 2001 and 
references therein). On long time scales, it displays clear 
luminosity--related spectral changes. This is illustrated in the 
observations reported by Barret \& Olive (2002; hereafter Paper I) 
who followed the spectral evolution of the source during a transition 
between soft and hard X-ray spectral states. The data were 
interpreted in the framework of a truncated accretion disk geometry 
for which the spectral transitions observed were associated with 
changes in the disk truncation radius.

In this paper, we follow the multi-Lorentzian approach to study the
rapid X-ray variability of the source using the same set of
observations as presented in Paper I. The aim of this work is to look
for correlations between the timing and spectral parameters to get
further insights on the accretion flow changes associated with the
state transitions.
 
\section{Observation and data analysis}

\begin{figure}[!t]
\begin{center}
\epsscale{1.0}
\plotone{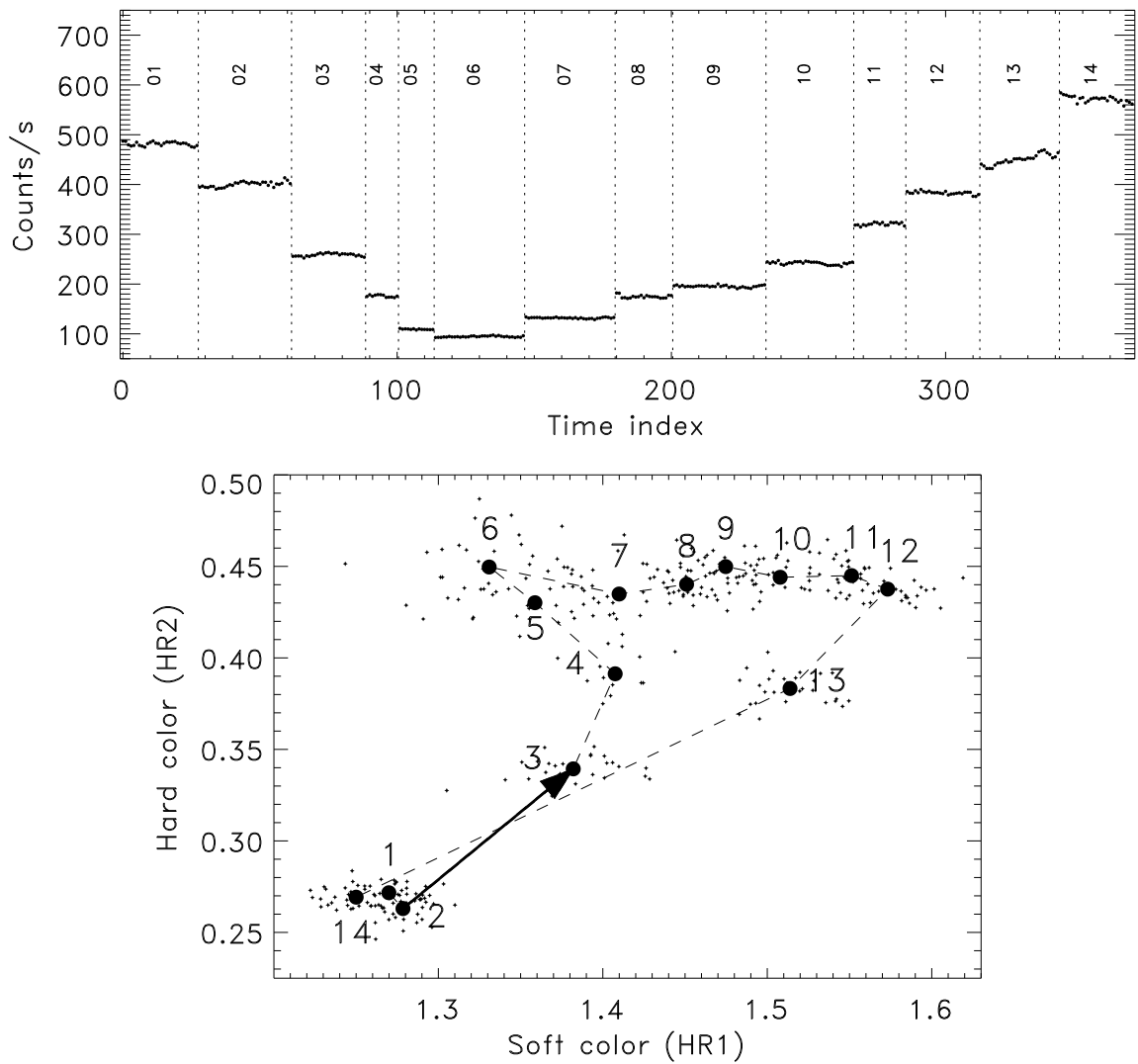}
\caption{The PCA background subtracted light curve of
  the \fouru~in the 3-16 keV band (top panel). The data are plotted
  versus a time index to avoid the gaps in time between observations. The bottom panel shows the X-ray color-color diagram of \fouru. The soft color (HR1) is
defined as the ratio between the 4.3-6.5 keV counts and 2.9-4.3 keV
counts, and the hard color (HR2) as the ratio between the 10.1-16.2
keV counts and 6.5-10.1 keV counts. The observation numbers allow to follow the path of the source during the state transitions. 
  \label{fig1_ogb}}
\end{center}
\end{figure}

We use the same set of \rxte~observations as in Paper I. The data
cover a period from February 10th to March 9th, 1999, with pointings
performed every two days, giving 14 observations in total. In the
following the observations are numbered from 1 to 14. During this
period, the source X-ray count rate first decreased, reached a minimum
around observation 6, and then increased to reach the initial level
(see top of Fig.\ \ref{fig1_ogb}). On the X-ray color-color diagram
(bottom of Fig.\ \ref{fig1_ogb}), the source traced out part of a more
or less triangular pattern (see also Gierli{\'n}ski \& Done
2002). \fouru~ evolved from the soft (the so-called banana) to the
hard (island) state in $\sim 10$ days, remained in a hard state for
$\sim 12$ days, and then moved back to the soft state in $\sim 4$
days. In Paper I, all fourteen energy spectra were adequately fitted
by the sum of a blackbody, thermal Comptonization and an iron line.

For timing analysis we use the science event data obtained  in 
configuration E\_16us\_64M\_0\_1s. We have computed 0.03-4096 Hz PDS 
in the 3--30 keV energy range, for the same time intervals as the 
energy spectra. The Poisson counting noise level was estimated from 
the data above 1500 Hz and subtracted from the PDS before fitting. 
Each PDS has been rebinned using a logarithmic scheme, and expressed 
in terms of fractional RMS amplitude. 

We have fitted each of the 14 PDS by up to 4 Lorentzians with the
following analytical formula given below (van Straaten et al.\ 2002):
\begin{equation} P_\nu = {r^2 \Delta \over \left({\pi
\over 2} + \arctan {2 Q}\right) \left[\Delta^2 + (\nu - 2 \Delta 
Q)^2\right]}, \end{equation} where $\Delta \equiv \nu_{\rm 
max}/\sqrt[]{1 + 4 Q^2}$ is the half width at half maximum of the 
Lorentzian and $\nu_{\rm max}$ (the position of the maximum of the 
Lorentzian per logarithmic frequency interval, i.e.\ the peak 
position of $\nu P_\nu$), $r$ (the fractional RMS integrated from 0 
to $\infty$) and $Q$ (the quality factor) are the independent 
parameters to fit. The centroid of the Lorentzian ($\nu_{0}$) is 
related to the above parameters as: $\nu_{0} \equiv 2\Delta 
Q=\sqrt[]{\nu_{\rm max}-\Delta^2}$. The spectral fitting and error 
computations were carried out using XSPEC 11.1 (Arnaud et al.\ 1996). 
Whenever the Q parameter was consistent with zero, we used 
zero-centered Lorentzians ($Q$ frozen at 0). Finally, we searched for 
kHz QPOs using the PDS scanning technique described in Boirin et 
al. (2000).

\section{Results}

\begin{figure}[!t]
\begin{center}
\epsscale{1.0}
\plotone{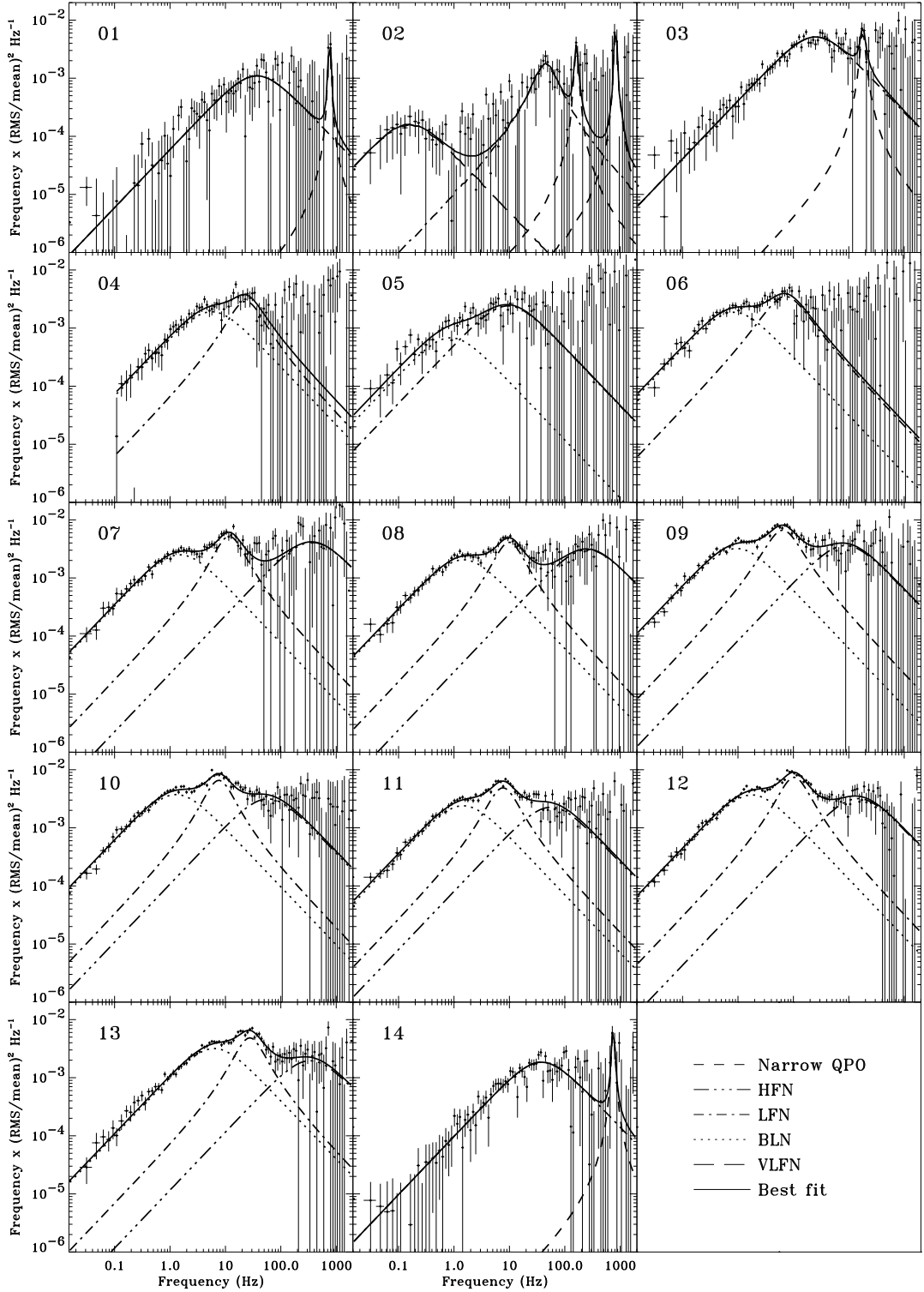}
\vspace*{-2cm}
\caption{The 14 power density spectra (PDS) from 
\fouru~fitted by up to four Lorentzians. HFN stands for High 
Frequency Noise, LFN for Low-Frequency Noise, BLN for Band-Limited 
Noise and VLFN for Very-Low Frequency Noise. Narrow Quasi-Periodic 
Oscillations (QPOs) were also detected in observations 1, 2, 3 and 
14.  
\label{fig2_ogb}}
\end{center}
\end{figure}

\begin{figure}[!t]
\begin{center}
\epsscale{1.0}
\plotone{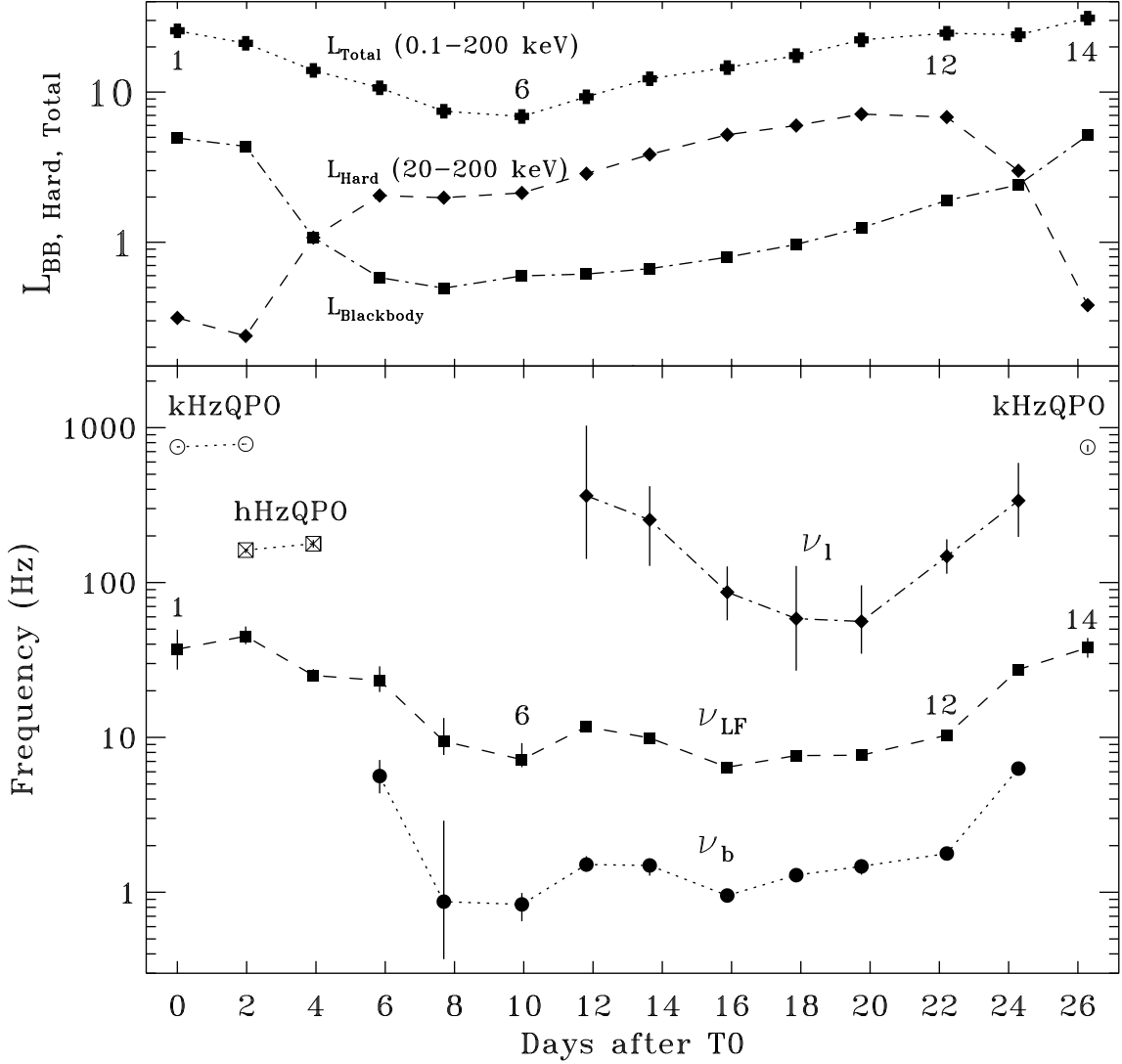}
\caption{Top panel: The luminosities ($L_{Total}$, $L_{Hard}$ and 
$L_{Blackbody}$) along the 14 observations in units of 
$10^{36}$\ergs. Bottom panel: Time evolution of the PDS fitted 
parameters. $\nu_b$ is the frequency of the  lorentzian fitting the 
BLN component of Fig.\ \ref{fig2_ogb}, $\nu_{\rm LF}$ is the 
frequency of the LF lorentzian, $\nu_l$ is the HFN Lorentzian. The 
naming convention follows Belloni et al. (2002). Hecto-Hz QPOs were 
detected in observations 2 and 3. kHz QPOs were detected in 
observations 1, 2 and 14. As can be seen  $\nu_{\rm LF}$ is the only 
component which is detected in all observations. \label{fig3_ogb}}
\end{center}
\end{figure}

The spectra together with the best-fitting models are presented in 
Fig.\ \ref{fig2_ogb}. The best fit parameters are listed in Tab.\ 
\ref{table1}. Time evolution of the fitted Lorentzian frequencies is 
shown in Fig.\ \ref{fig3_ogb}. In those observations with the best
statistics (7 to 13), three broad components are detected with noise
extending up to $\ge 500$ Hz. kHz QPOs around 750--800 Hz were
detected above the 3$\sigma$ level in observations 1, 2 and 14 (see
Tab.\ \ref{table2}) when the source was in the soft state.  Our
frequencies are close to those reported by Ford et al.\ (1998) using
observations performed in February to June 1997 while the source was
in a similar spectral state. In terms of fractional RMS, the
sensitivity to QPO detection decreases with the source count rate. For
indication, 3$\sigma$ upper limits of $4.6, 8.5, 6.2, 5.0, 4.5, 3.4
$\% can be derived for a QPO at 750 Hz with a 20Hz width in
observations 3, 5, 7, 9, 11, 13 respectively. However, in terms of
modulated amplitude, the signals detected in observations 1, 2 and 14
would have been clearly detected in the other observations.

Narrow QPOs around $\sim 170$ Hz were found in observations 2 and 3 
($\nu=162^{+3}_{-7}$ Hz, RMS(\%)=$2.9\pm0.7$ for observation 2 and 
$\nu=178^{+10}_{-10}$ Hz, RMS(\%)=$4.5\pm1.0$ for observation 3). 
Similar QPOs, called hecto-Hz QPOs have been reported from other 
sources (e.g. van Straaten et al.\ 2002). The PDS of observation 2 is 
also remarkable by the presence of a very low frequency component 
(VLFN, see Fig.\ \ref{fig2_ogb}), not seen in other observations. A 
fit with a zero-centered Lorentzian of this component yields 
$\nu_{\rm max} = 0.156_{-0.04}^{+0.06}$ Hz and RMS(\%) = $3.1\pm0.4$. 
These transient features will not be discussed in this paper.

\subsection{Identification of the PDS components}

\begin{figure}[!t]
\begin{center}
\epsscale{1.0}
\plotone{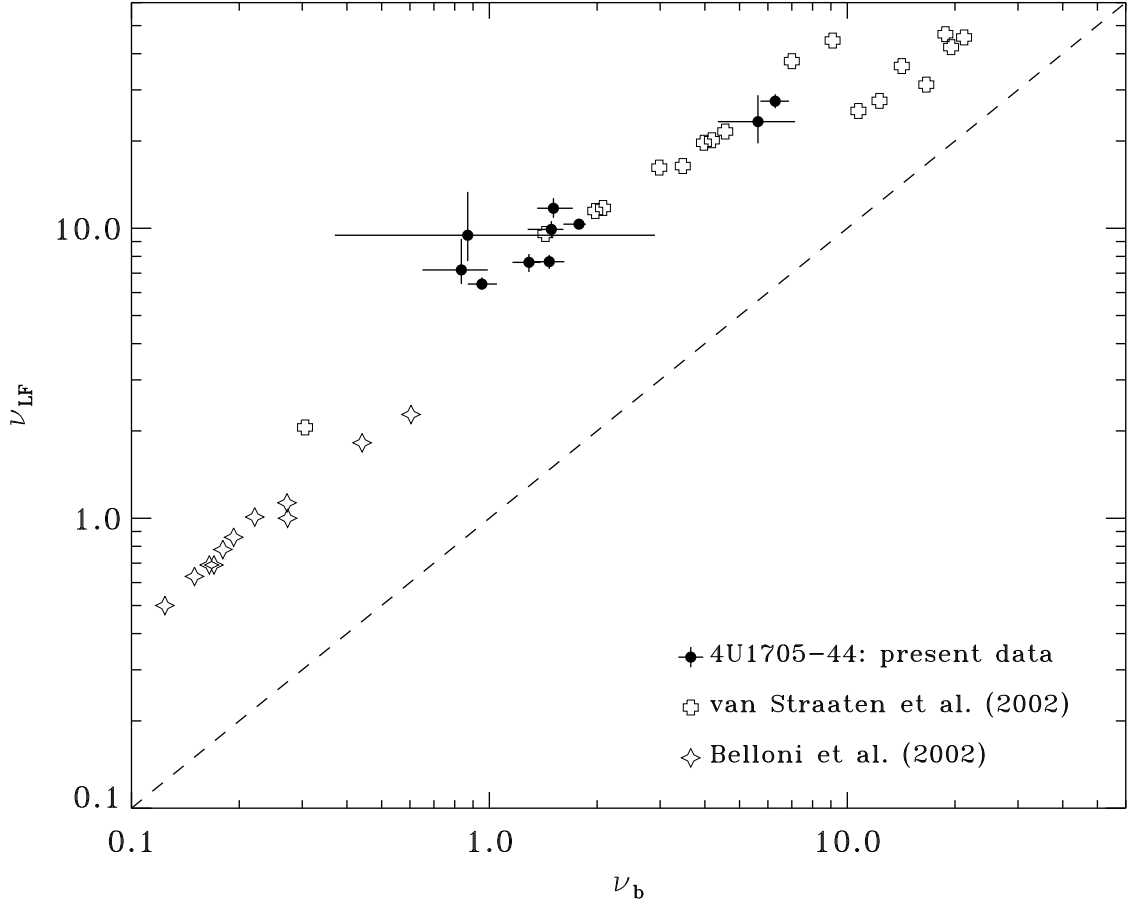}
\caption{\nulf~as a function of \nub. Data points from \fouru~are 
plotted with a filled circle. They are compared with those obtained 
for systems similar to \fouru~for which the same type of 
multi-lorentzian fitting of the PDS was performed (Empty stars : 
GS~1826--34, 1E~1724--3045 and SLX~1735--369; Belloni et al. (2002). 
Empty crosses : 4U~0614+091 and 4U~1728--34; van Straaten et al. 
(2002)) Our data point fit in the correlation, indicating that the 
identification of the various frequencies of the PDS is correct. 
\label{fig4_ogb}}
\end{center}
\end{figure}

The PDS of \fouru~resemble those reported by Belloni et al.\ (2002) 
and van Straaten et al.\ (2002)  from similar systems.  Following 
Belloni et al.\ (2002), we name the first two Lorentzians  as band 
limited noise (BLN at $\nu_{b} \sim$ 1--7 Hz) and low-frequency 
noise (LFN at $\nu_{\rm LF} \sim$ 7--40 Hz). The LFN component is the 
only one which can be identified in all observations. In Fig.\ 
\ref{fig4_ogb}, we plot $\nu_{\rm LF}$ versus $\nu_{b}$. A clear 
correlation can be seen. Our results can be directly compared with 
those obtained for some other neutron star LMXBs. Taking out the same 
two components in Belloni et al.\ (2002) for 1E~1724-3045 and 
GS~1826-34 and van Straaten et al.\ (2002) for 4U~1728-34 and 
4U~0614+091 (in the latter paper $\nu_{b}$ is called $\nu_{\rm 
BLN}$), we find that our data points fall in on the global 
correlation previously reported by these authors. This supports the 
identification of the first two components of the PDS of \fouru~with 
BLN and LFN.

\begin{figure}[!t]
\begin{center}
\epsscale{1.0}
\plotone{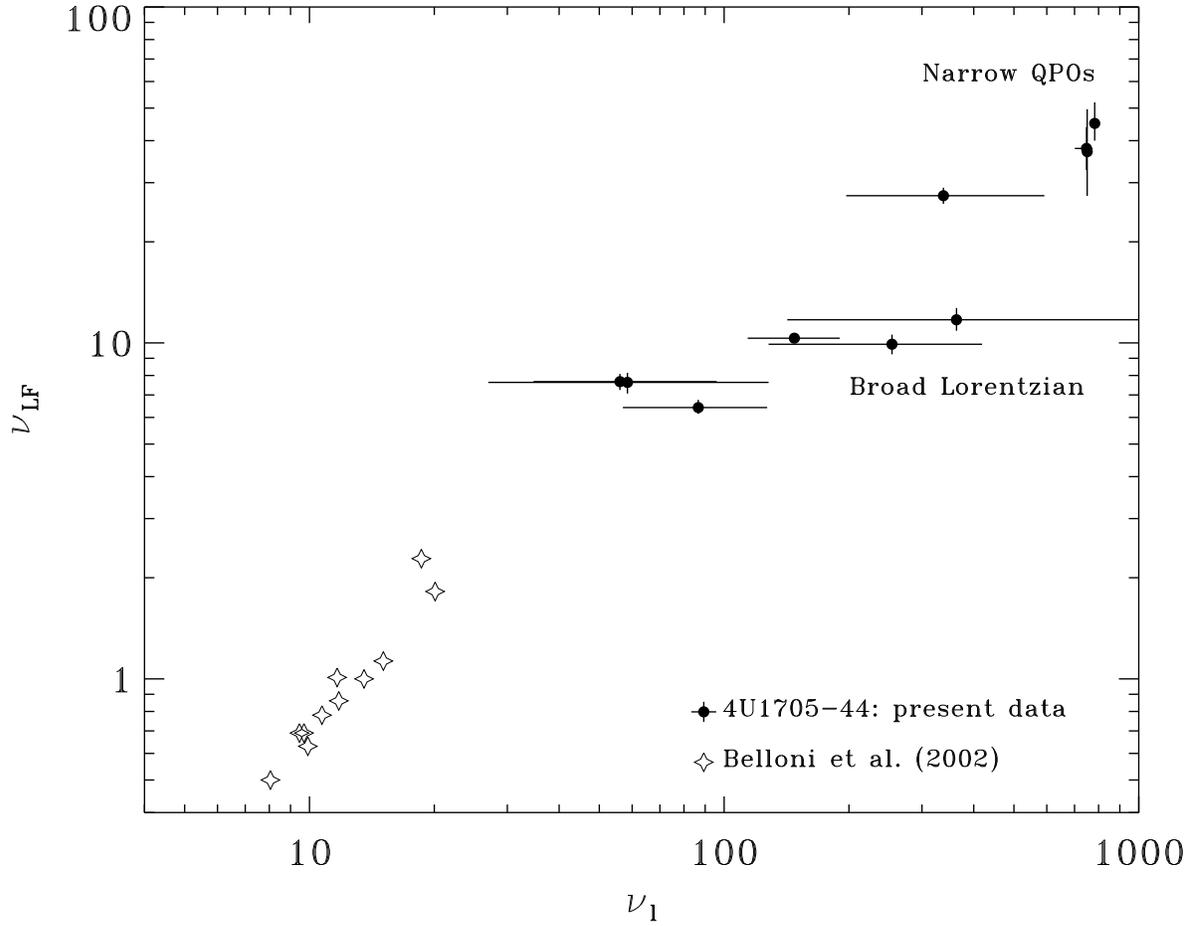}
\caption{\nulf~as a function of \nul. For comparison data points from 
Belloni et al. (2002) are also shown. Again, a correlation has been 
shown to exist in many systems (Psaltis et al. 1999; Belloni et al. 
2002). \nul~and the three kHz QPO frequencies follow the 
correlation. \label{fig5_ogb}}
\end{center}
\end{figure}

The third broad Lorentzian component which we denote as high 
frequency noise (HFN at characteristic frequencies of $\sim$ 60--400 
Hz) is detected only in data sets 7 to 13. It is the one which is the 
less statistically significant, and its identification is therefore 
less certain. It may be the hecto-Hz component reported by van 
Straaten et al. (2002) from similar systems. However, these hecto-Hz 
components are generally narrower than the features we observed in 
\fouru~which are fitted with zero-centered Lorentzians. It seems 
therefore more likely to be what is called $L_l$ in Belloni et al.\ 
(2002); namely the extrapolation towards low frequencies of the lower 
kHz QPO. Support from that hypothesis comes from the fact our data 
points match the global correlation between $\nu_{\rm LF}$ and 
$\nu_l$ discovered by Psaltis et al.\ (1999). This feature would then 
be closely related to the lower kHz QPOs detected in observations 1, 
2 and 14. As expected, the kHz QPO frequencies fit in the global 
correlation (see Fig.\ \ref{fig5_ogb}).

\subsection{Correlation between PDS and energy spectrum parameters}

Since the LFN component is detected in all observations and since the 
other frequencies are correlated with $\nu_{\rm LF}$, we used 
$\nu_{\rm LF}$ for searching correlations between PDS and energy 
spectrum parameters. 

\begin{figure}[!t]
\begin{center}
\epsscale{1.0}
\plotone{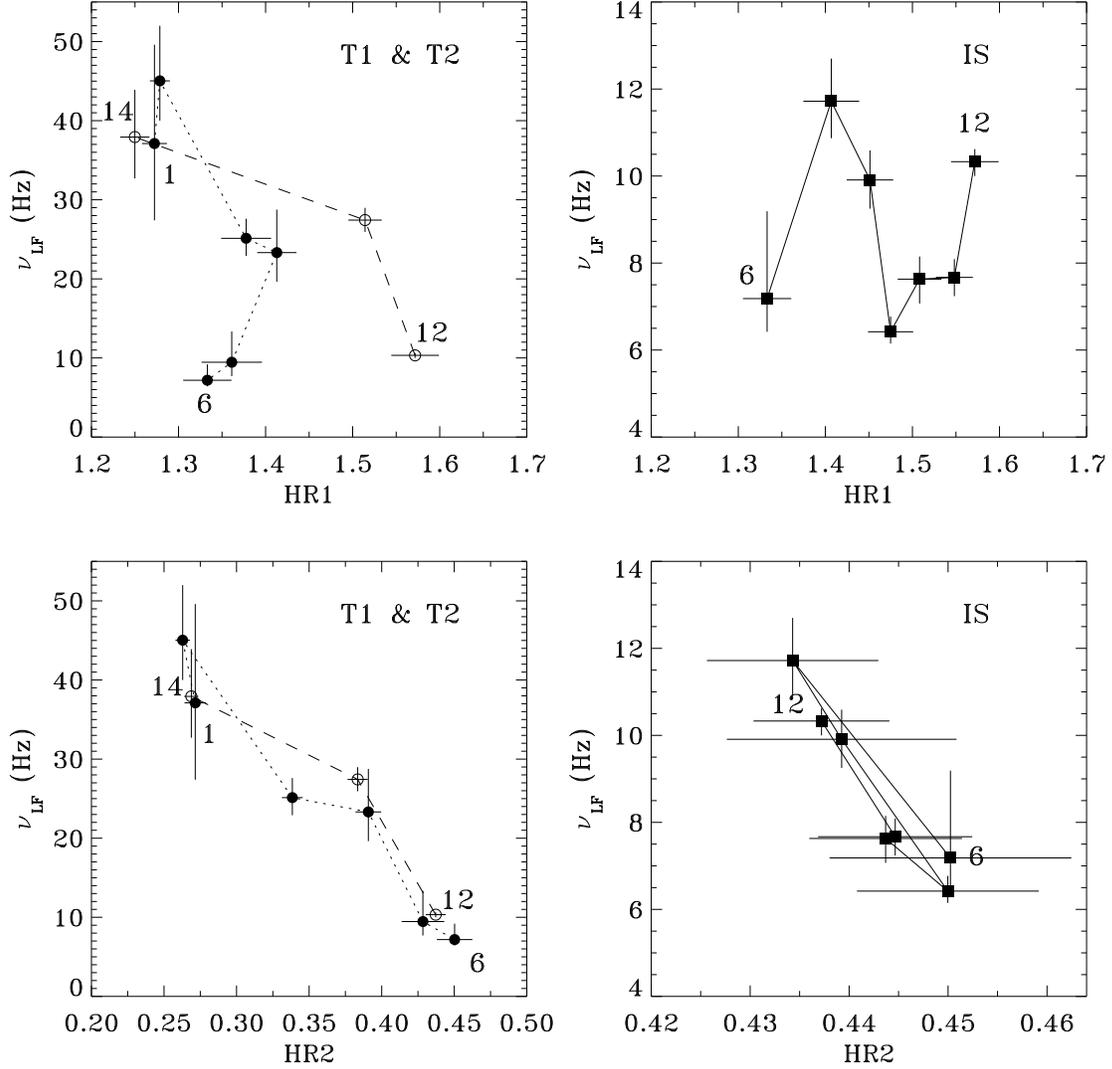}
\caption{\nulf~as a function of HR1 and HR2. Three segments of the 
observations are considered: T1 is the transition which occurred 
between observations 1 and 6 (filled circles), IS is the Island State 
between observations 6 and 12 (empty squares) and T2 is the 
transition between observations 12 and 14 (empty circles). 
\label{fig6_ogb}}
\end{center}
\end{figure}

\begin{figure}[!t]
\begin{center}
\epsscale{1.0}
\plotone{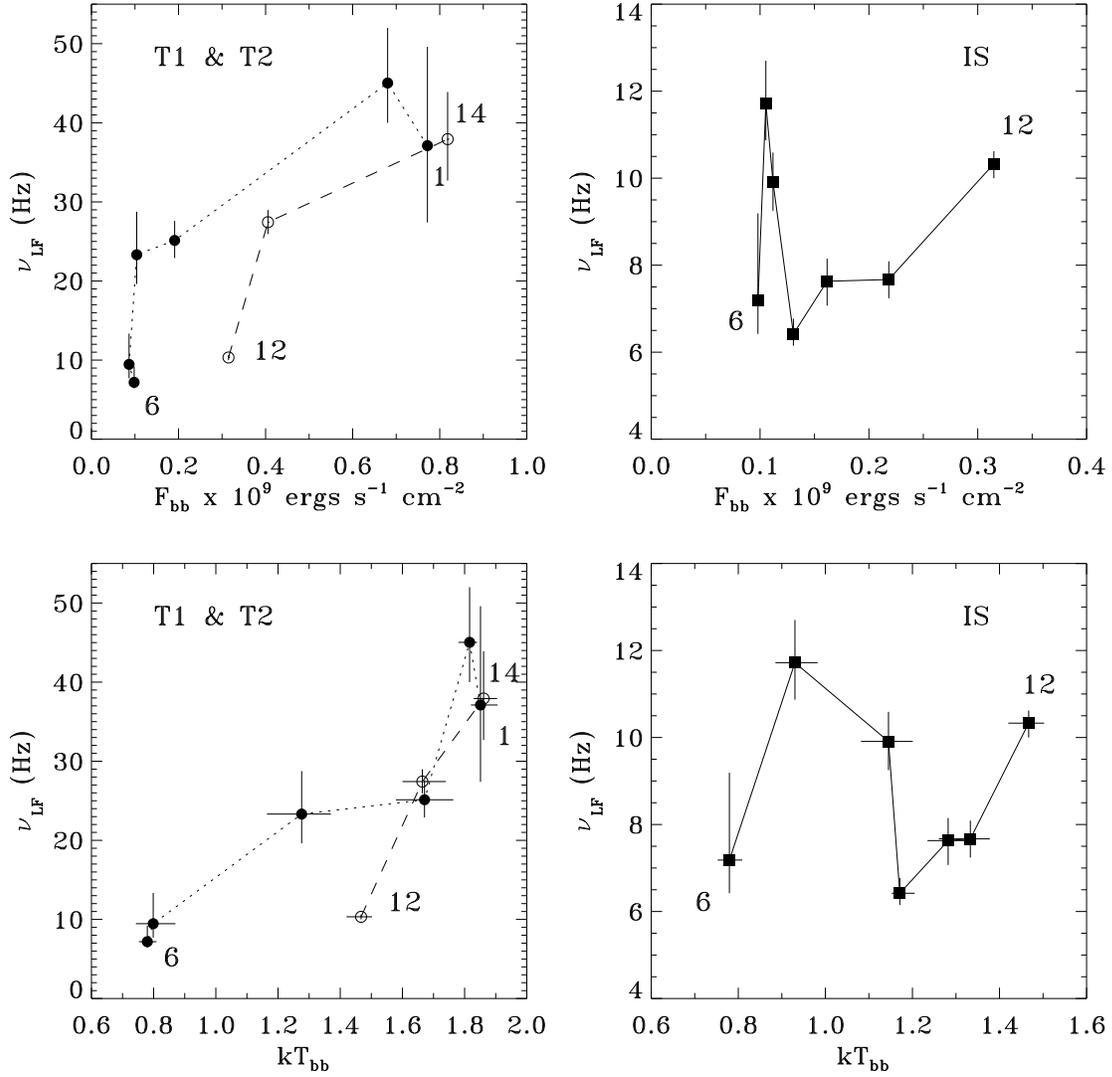}
\caption{\nulf~as a function of the blackbody temperature and flux 
for the three segments of observations.
\label{fig7_ogb}}
\end{center}
\end{figure}

\begin{figure}[!t]
\begin{center}
\epsscale{1.0}
\plotone{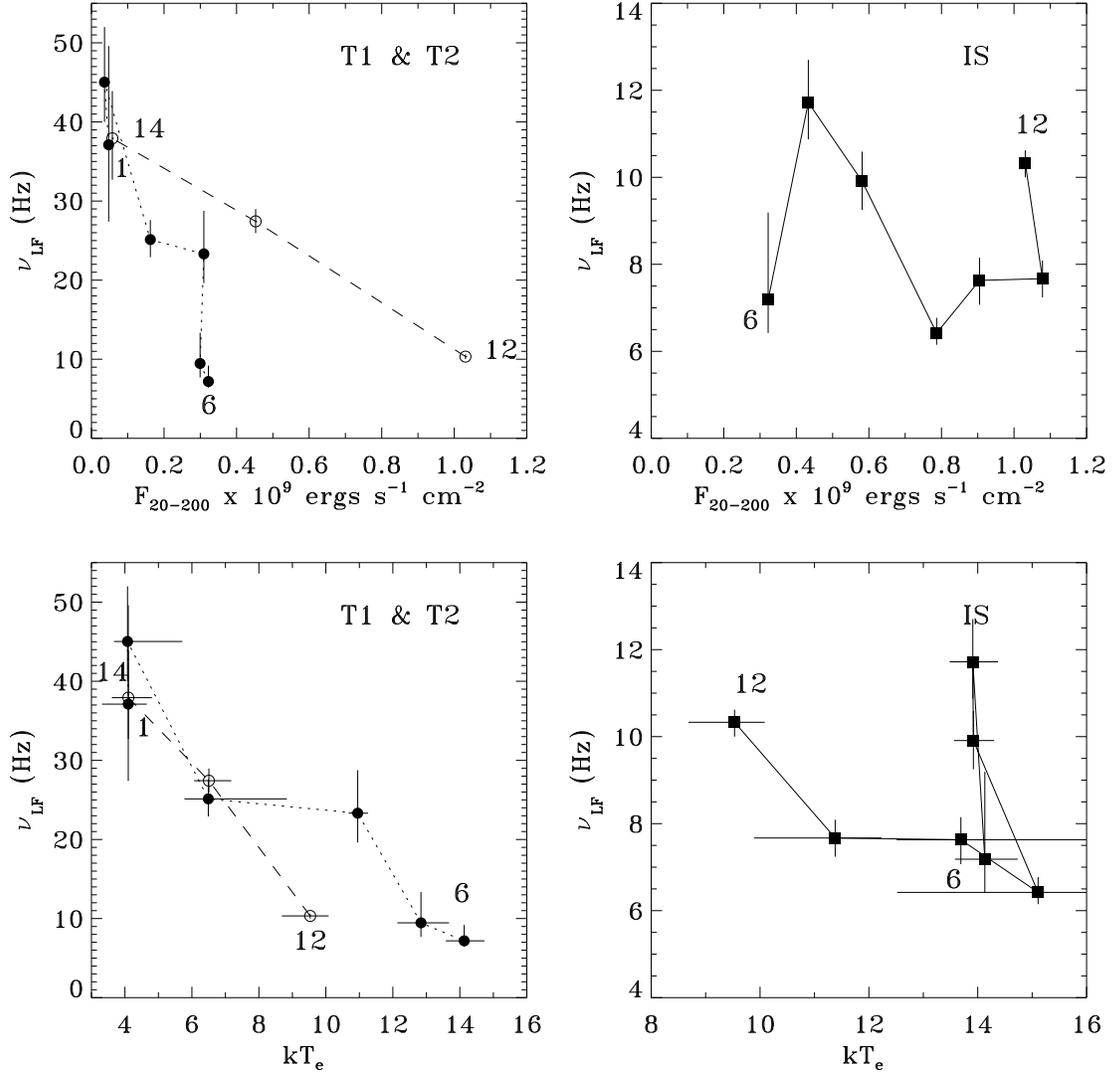}
\caption{\nulf~as a function of the electron temperature and the 
bolometric flux between 20 and 200 keV for the three segments of 
observations.
\label{fig8_ogb}}
\end{center}
\end{figure}

We split the observation in 3 segments: the transition the soft to 
hard state (T1, obs.\ 1 to 6), the observations corresponding to the 
island state (IS, obs.\ 6 to 12), the transition hard to soft (T2, 
obs.\ 12 to 14). In Fig.\ \ref{fig6_ogb},  \ref{fig7_ogb} and 
\ref{fig8_ogb}, we plot $\nu_{\rm LF}$ versus the soft and hard 
colors, $\nu_{\rm LF}$ versus the blackbody temperature and 
bolometric flux and $\nu_{\rm LF}$ versus the electron temperature of 
the Comptonizing cloud  (\kte)  and  20--200 keV hard X-ray flux 
(\fhard). 

Fig.\ \ref{fig6_ogb} shows that there is a clear anti-correlation
between \nulf~and HR2 in all three segments. On the other hand, there
are no obvious correlations between \nulf~and HR1.  This is consistent
with previous work which have shown that timing features, such as
kHz QPOs (whose frequencies correlate with \nulf, see above)
anti-correlate with the hard color (e.g. M\'endez et al.\ 1999).

Fig.\ \ref{fig7_ogb} clearly shows a nice correlation between 
$\nu_{\rm LF}$ and the blackbody temperature and flux at 
the start of the soft to hard transition (obs.\ 1 to 3) and during 
the reverse transition (obs.\ 12 to 14). Similar correlations between 
the blackbody flux and kilo-QPO frequencies have been already 
reported from 4U~0614+091 (Ford et al.\ 1997). Here we show that the 
correlations break down in 4U~1705--44 when the blackbody flux dropped 
below $\sim 10^{-10}$ \ergscm. Down in the IS, the absence of 
correlation is striking, despite  the smooth but significant increase 
of the blackbody flux and temperature.

Finally, Fig.\ \ref{fig8_ogb} shows that during T1 and T2  there is a 
clear anti-correlation between  \nulf~and both \kte~and \fhard, and 
there are no obvious correlations of these two parameters with 
\nulf~in IS (no correlations with time are present either). 

\section{Discussion} 

In most of the rapid variability models the frequencies of particular
noise components are set at some transition radius in the disk (e.g.
Stella et al.\ 1999; Psaltis \& Norman 2002). It could be for 
instance
a truncation of a standard Shakura-Sunyaev disk into a hot inner
flow. Here we assume that the frequencies we observe, in particular
$\nu_{\rm LF}$ which can be traced throughout all the data sets, are
related to the truncation radius of the disk. 
In Paper I we interpreted the spectral state transitions in the
framework of a such truncated accretion disk geometry, allowing the
truncation radius to change within the observations. Let us now
discuss the new constraints that the data presented above set on this
scenario.

At the beginning of our observations (obs.\ 1--2), the source is in 
the soft state, the truncation radius is small and the disk is close 
to the neutron star surface (see Paper I). The characteristic 
frequency \nulf~is at its maximum around 40 Hz, which is consistent 
with the inner disc radius reaching its minimum. It is also the time 
during which the kHz QPOs at 750--780 Hz are detected. If the latter 
one is associated with the Keplerian frequency at the truncation 
radius, the disk truncates at $\sim 20$ km. The soft component 
($kT_{\rm BB} \sim 1.8$ keV) present in the energy spectrum 
originates from the disk (Paper 1). The hard component is from the 
boundary layer, with low temperature of $kT_e \sim 4$ keV.

In Paper I, we suggested that during the transition to the hard state
(T1, obs.\ 2 to 6) the disk receded while the total source luminosity
decreased from $\sim 2.1$ to $0.7 \times 10^{37}$\ergs.  This is
supported by our timing analysis: during the transition \nulf~dropped
roughly by factor of $\sim 6$. The fastest variability timescale
around a compact object is given by the Keplerian orbital frequency
(at $\nu_k$). This sets an upper bound on the inner radius of the disk
at which the variability is produced (see e.g. di Matteo \& Psaltis
1999). When $\nu_k$ is not detected, one can use the empirical
correlation found by Psaltis et al. (1999) between $\nu_k$ and
\nulf~to estimate the inner disk radius as a function of \nulf~(see
equation 2 in di Matteo \& Psaltis 1999). Using the latter equation,
one obtains a disk truncation radius around 70 km for observation
6. When the disk recedes, the temperature and flux of the soft
component drop to their lowest values (\ktbb $\sim 0.8$ keV, Fig.\
\ref{fig7_ogb}, left panels).  Meanwhile, the hot inner flow (or
optically thin boundary layer) builds up and its temperature rises
from $\sim$ 4 to $\sim$ 14 keV (Fig.\ \ref{fig8_ogb}, left panels)
possibly because the disk contribution to the cooling of the inner
flow is then seriously reduced.

Between observations 6 and 12 (IS) the source is in the hard state 
while the mass accretion rate (inferred from the source luminosity 
$L_{0.1-200} \sim 0.7$ to $2.5 \times 10^{37}$\ergs) 
monotonically increased. The energy spectrum is dominated by the 
emission from the hot optically thin inner flow (Paper I). In the 
proposed scenario, the disk is far away, and \nulf~is low, $\sim$ 
7--12 Hz. With the exception of observation 6, all the PDS show three 
broad Lorentzians (at \nub, \nulf~and \nul) at characteristic 
frequencies which are roughly constant suggesting that the geometry 
of accretion remains similar through these observations. There is no 
correlation between \nulf~(which is roughly stable) and the blackbody 
flux and temperature either (Fig.\ \ref{fig7_ogb}, right panels) 
despite significant variations of these parameters. During this hard 
state, the spectral and the timing parameters are {\em 
de-correlated\/}. This can be naturally explained if the soft 
component (blackbody) originates from the neutron star surface, not 
from the disk. Basically, the disk is far away and its temperature is 
so low that it cannot be seen by the PCA, so it does not contribute 
to the energy spectrum. We can see its variations only through the 
timing properties, as \nulf. The blackbody comes from the neutron 
star surface which is continuously heated along these observations. 
This soft component is further Comptonized in the optically thin hot 
inner flow/boundary layer leading to an increasing $F_{20-200}$ 
flux (Fig.\ \ref{fig8_ogb}, right panel). It is also important to 
note that at the end of the IS state (obs.\ 12), the source has 
recovered an accretion rate (as inferred with the luminosity) similar 
to observations 1 or 2, while the truncation radius of the disk (as 
traced with \nulf) is still large.

The transition back to the soft state (T2, obs.\ 12 to 14) is the
inverse of the T1 transition. The blackbody flux and temperature
increased, which was interpreted (Paper I) as the reappearance in the
PCA energy band of the disc moving in and becoming hotter. Again, the
timing analysis supports this result: there is a significant increase
in \nulf, very well correlated both with $T_{\rm bb}$ and $F_{\rm
bb}$. Meanwhile, the inner flow is significantly cooled by the
increasing flux of soft photons coming from the disk and its
temperature falls from $\sim$ 10 to $\sim$ 4 keV. In observation 14,
the system is back to a state very similar to observation 1 and 2, and
we start seeing again kHz QPO at $\sim 750$ Hz.

It is also interesting to look at rapid X-ray variability as a 
function of the position in the color-color diagram. Both atolls and 
Z sources trace out a characteristic track in the diagram (see e.g.\ 
Hasinger \& van der Klis 1989). The frequencies in the PDS are 
usually very well correlated with the position in the diagram. It has 
been shown recently that at low luminosities atolls form an 
additional, upper horizontal branch in the color-color diagram, 
tracing out a Z-shaped track, similar to Z sources (Gierli{\'n}ski \& 
Done 2002; Muno, Remillard \& Chakrabarty 2002).

\fouru~moved from left to right in the horizontal branch between
observations 6 and 12, and then along the diagonal, between
observations 12 and 14 (Fig.\ \ref{fig1_ogb}). The timing and spectral
properties on the diagonal branch are typical of other atoll
sources. However, the upper branch is different: as we discussed
previously, there is no correlation between \nulf~and the position on
the branch, which in this case (branch is horizontal) can be described
by the soft color HR1 (see Fig.\ \ref{fig6_ogb}).

In our scenario, at low accretion rate, the position in the upper 
branch of the diagram is set by the neutron star temperature, which 
also supplies the seed photons. With the increasing accretion rate, 
$\dot{M}$, this temperature increases, so does the temperature of the 
low-energy cutoff in the Comptonized component. The spectrum below 4 
keV hardens, so the soft color increases and the source moves to the 
right in the diagram. This movement is {\em not\/} accompanied by the 
decrease of the truncation radius in the disk. Then, on the diagonal 
branch the disk temperature increases, so it can be seen in the PCA 
spectrum, while the neutron star surface is obscured by the 
increasing optical depth of the boundary layer. Thus, both spectral 
and timing properties depend now on the disk, and therefore are 
correlated. In this picture, the position on the upper branch depends 
on $\dot{M}$, while the characteristic frequency \nulf~do not.

The path of \fouru~in the color-color diagram is even more
complicated. On its way from the banana to the island state (T1) it
traces out a different track than in the opposite direction (T2). The
timing and spectral properties on both tracks are different. For
instance observations 3 and 13 have very similar \nulf, but different
bolometric luminosities (L$_{\rm 0.1-200}=1.4 \times
10^{37}$\ergs, $2.4 \times 10^{37}$\ergs~respectively, see Paper I)
and different power density spectra (see Fig.\ \ref{fig2_ogb}). This
shows that for the same truncation radius, one can have different
instantaneous accretion rate as derived from the source luminosity,
and different time variability (the Lorentzians at \nub~and \nul~are
not detected in observation 3). In the proposed scenario, there are
however important differences between observations 3 and 13 (and more
generally between T1 and T2). In the first case, the disk recedes, the
luminosity is decreasing, whereas in obs.\ 13 the disk approaches the
neutron star and the luminosity is increasing. This means that the
state of the system depends at some level on its recent history. Any
models proposed to explain the changes in \nulf~must take this fact
into account.

One possibility could be that \nulf~(i.e the truncation radius) is 
set by some long-time averaging process over time scales of days and 
only depends on a parameter $\eta = 
\dot{M_{d}}/<\!\!\dot{M_{d}}\!\!>$ (where $\dot{M_{d}}$ is the 
accretion rate through the disk) similarly to what has been proposed 
by van der Klis (2001) to explain the parallel tracks phenomenon in 
LMXBs. This model naturally predicts the decorrelation between the 
truncation radius position and the total accretion rate (and so the 
luminosity). In that scheme, the motion of the disk is associated 
with a rapid decrease of $\dot{M_{d}}$ during the T1 transition (i.e 
$\eta < 1$, decreasing with time, the disk recedes) and a reversed 
evolution during T2. The disk is stable either if $\dot{M_{d}}$ is 
constant or if it varies linearly with time over several days. This 
could be the case from observations 6 to 12 when \nulf~was low and 
roughly stable.

\section{Conclusions}

The spectral and timing data of the state transitions observed from 
\fouru~can be explained in the framework of a model, in which the 
critical parameter is the position of the truncation radius between 
the disk and a hotter inner flow. This parameter sets both the 
frequencies of the timing features and the spectral shape. What sets 
the value of the truncation radius is unclear at the moment, but both 
the timing and spectral data indicate that it cannot be the mass 
accretion rate derived from the bolometric source luminosity. One 
possibility could be this radius is set by some long-time averaging 
process over time scales of days.

\clearpage

\section{Tables}
\begin{table*}[!h]
\begin{center}
    \begin{tabular}{cllllllll}
\hline
 Obs &  $\nu_{b}$ & RMS$_{\rm b}$ & $\nu_{\rm LF}$ & RMS$_{\rm 
 LF}$ & $Q_{\rm LF}$ & $\nu_{l}$ & RMS$_{\rm l}$ & $\chi^2/d.o.f$ 
 \\
\hline
01 & \ldots & \ldots & $37.11^{+12.5}_{-9.7}$&$5.9^{+0.6}_{-0.6}$&(0) & \ldots & \ldots & 101.3/93 \\
02 & \ldots & \ldots & $45.03^{+7.0}_{-5.0}$&$4.9^{+0.9}_{-0.9}$&$0.91^{+0.8}_{-0.5}$ & \ldots & \ldots & 77.3/89 \\
03 & \ldots & \ldots & $25.13^{+2.5}_{-2.2}$&$12.7^{+0.4}_{-0.5}$&(0) & \ldots & \ldots & 96.8/94 \\
04 & $5.63^{+1.5}_{-1.3}$&$7.8^{+1.0}_{-1.4}$ & $23.32^{+5.4}_{-3.7}$&$7.3^{+3.0}_{-1.8}$&$0.56^{+0.5}_{-0.4}$ & \ldots & \ldots & 105.8/87 \\
05 & $0.87^{+2.0}_{-0.5}$&$4.6^{+0.4}_{-0.4}$ & $9.46^{+3.9}_{-1.8}$&$8.7^{+0.8}_{-2.2}$&(0) & \ldots & \ldots & 102.1/93 \\
06 & $0.83^{+0.2}_{-0.2}$&$7.7^{+0.2}_{-0.3}$ & $7.18^{+2.0}_{-0.8}$&$8.9^{+3.5}_{-0.6}$&$0.36^{+0.1}_{-0.4}$ & \ldots & \ldots & 98.2/93 \\
07 & $1.51^{+0.2}_{-0.1}$&$8.9^{+0.4}_{-0.4}$ & $11.72^{+1.0}_{-0.9}$&$9.3^{+1.1}_{-1.1}$&$0.67^{+0.2}_{-0.2}$ & $363^{+667}_{-221}$&$11.3^{+2.8}_{-2.8}$ & 98.1/91 \\
08 & $1.49^{+0.1}_{-0.2}$&$8.0^{+0.3}_{-0.4}$ & $9.91^{+0.7}_{-0.7}$&$8.2^{+1.0}_{-0.8}$&$0.67^{+0.2}_{-0.2}$ & $254^{+165}_{-126}$&$9.8^{+1.5}_{-1.6}$ & 102.3/91 \\
09 & $0.95^{+0.1}_{-0.1}$&$10.0^{+0.2}_{-0.2}$ & $6.42^{+0.3}_{-0.3}$&$11.1^{+0.9}_{-0.5}$&$0.55^{+0.1}_{-0.1}$ & $86^{+40}_{-29}$&$10.6^{+8.1}_{-0.8}$ & 127.6/91 \\
10 & $1.29^{+0.1}_{-0.1}$&$10.8^{+0.3}_{-0.4}$ & $7.63^{+0.5}_{-0.6}$&$10.3^{+1.6}_{-1.9}$&$0.69^{+0.3}_{-0.2}$ & $58^{+69}_{-31}$&$9.9^{+0.7}_{-0.7}$ & 127.0/91 \\
11 & $1.47^{+0.1}_{-0.2}$&$8.7^{+0.4}_{-0.4}$ & $7.67^{+0.4}_{-0.4}$&$9.1^{+1.1}_{-1.1}$&$0.65^{+0.2}_{-0.1}$ & $56^{+39}_{-21}$&$8.4^{+0.6}_{-0.7}$ & 112.9/91 \\
12 & $1.78^{+0.1}_{-0.2}$&$10.7^{+0.2}_{-0.4}$ & $10.33^{+0.3}_{-0.3}$&$11.2^{+0.7}_{-0.5}$&$0.65^{+0.1}_{-0.1}$ & $147^{+42}_{-33}$&$10.0^{+0.7}_{-0.7}$ & 149.0/91 \\
13 & $6.29^{+0.6}_{-0.6}$&$10.0^{+0.4}_{-0.5}$ & $27.43^{+1.5}_{-1.5}$&$8.8^{+0.9}_{-0.9}$&$0.69^{+0.2}_{-0.1}$ & $338^{+254}_{-141}$&$7.8^{+1.1}_{-0.6}$ & 100.3/91 \\
14 & \ldots & \ldots & $37.93^{+6.0}_{-5.2}$&$7.6^{+0.4}_{-0.4}$&(0) & \ldots & \ldots & 104.6/93 \\
\hline
\end{tabular}
    \caption{Best fit parameters for the 14 PDS. \nub~and RMS$_{\rm b}$ are respectively the frequency and the RMS of the band-limited noise component fitted with a zero-centered lorentzian.  \nulf~RMS$_{\rm LF}$ and Q$_{\rm LF}$ are the frequency, the RMS and the Quality factor of the Low-Frequency Noise component detected in all observations. \nul~and RMS$_{\rm l}$ are the frequency and RMS of the high frequency noise componant. The naming convention follows Belloni et al. (2002). Finally, $\chi^2/d.o.f$ is the $\chi^2$ value and the number of degree of freedom of the fit.}
        \label{table1}
\end{center}
\end{table*}

\begin{table*}[!h]
\begin{center}
    \begin{tabular}{clll}
\hline
 Obs &  $\nu$ (Hz) & $RMS$ (\%) & $Q$ \\
\hline
01 & $751^{+6}_{-5}$ & $3.5^{+1.2}_{-0.8}$ & $31^{+33}_{-13}$ \\
02 & $783^{+8}_{-7}$ & $2.9^{+1.1}_{-1.0}$ & $41^{+40}_{-24}$ \\
14 & $748^{+26}_{-47}$ & $3.8^{+1.0}_{-1.0}$& (10) \\
\hline
\end{tabular}
    \caption{The best fit parameters (frequency, RMS and the quality
    factor Q) for the kHz QPOs detected in observations 1, 2 and 14.
    For observation 14, the Q value was poorly constrained and then
    frozen at 10 for the fit. These frequencies are consistent with
    those reported by Ford et al. (1998).}
    \label{table2}
\end{center}
\end{table*}

\end{document}